\documentclass[aps,prb,twocolumn]{revtex4}

\usepackage{bm}
\usepackage{graphicx}

\begin{document}

\title{Non-local Andreev reflection in superconducting quantum dots}
\author{Dmitri S. Golubev and Andrei D. Zaikin}
\affiliation{Forschungszentrum Karlsruhe, Institut f\"ur
Nanotechnologie, 76021, Karlsruhe, Germany} \affiliation{I.E. Tamm
Department of Theoretical Physics, P.N. Lebedev Physics Institute,
119991 Moscow, Russia}

\begin{abstract}

With the aid of the Keldysh technique we develop a  microscopic theory of  
non-local electron transport in three-terminal NSN structures consisting of a 
chaotic superconducting quantum dot attached to one superconducting and 
two normal electrodes. Our theory fully accounts for 
non-equilibrium effects and disorder in a superconducting
terminal. We go beyond perturbation theory in
tunneling and derive a general expression for the system conductance matrix 
which remains valid in both weak and strong tunneling limits.
We demonstrate that the proximity effect yields a decrease 
of crossed Andreev reflection (CAR). Beyond weak tunneling limit 
the contribution of CAR to the non-local conductance does not cancel 
that of direct electron transfer between two normal terminals. 
We argue that temperature dependence of the non-local resistance of 
NSN devices is determined by the two competing processes -- 
Andreev reflection and charge imbalance -- and it has a  
pronounced peak occurring at the crossover between these two processes. 
This behavior is in a good agreement with recent experimental observations.
\end{abstract}


\maketitle

\section{Introduction}

Non-local (crossed) Andreev reflection \cite{BF,DF} is the process
which occurs in multi-terminal hybrid normal
metal-superconductor-normal metal (NSN) proximity structures and
involves two subgap electrons entering a superconductor from two
{\it different} normal terminals and forming a Cooper pair there.
This is in contrast to the standard mechanism of (local) Andreev
reflection \cite{And} (AR) in which case two subgap electrons
enter a superconductor from the same normal electrode through the
same interface. The phenomenon of crossed Andreev reflection (CAR)
manifests itself, e.g., in the dependence of the current $I_L$
through the left NS interface of an NSN structure on the voltage
$V_R$ across the right NS interface. As a result, the non-local
conductance $G_{LR}=\partial I_L/\partial V_R$ of an NSN device
differs from zero and can be detected experimentally. Such
experiments have recently been performed by several groups
\cite{Beckmann,Teun,Venkat} providing a number of interesting
observations some of which remain not fully understood.

It is important to mention that CAR is not the only process which
contributes to the non-local conductance $G_{LR}$. Another
relevant process is direct electron transfer (DET) between two normal
terminals through the superconductor. In the tunneling limit this
process is nothing but the so-called elastic cotunneling (EC).
It turned out \cite{Falci} that in the lowest order in tunneling
the contributions from EC and CAR to $G_{LR}$ exactly cancel each other
in the limit of low temperatures and voltages, i.e. the non-local conductance
$G_{LR}$ should vanish in this limit.

Note that this result \cite{Falci} is applicable only provided
transmissions of both NS interfaces remain small which is not
always the case in the experiments. At higher transmissions
processes to all orders should be taken into account and the
contributions of DET and CAR do not anymore cancel each other.
Hence, $G_{LR}$ does not vanish beyond the tunneling limit. In the
case of ballistic electrodes a non-perturbative (in barrier
transmissions) theory was recently developed by Kalenkov and one
of the authors \cite{KZ06,KZ07}. This theory allowed to
study the non-local conductance of NSN devices at arbitrary
transmissions leading to a conclusion that CAR contribution to
$G_{LR}$ {\it vanishes} in the limit of fully open NS barriers.
This result might seem counterintuitive since ordinary (local) AR
reaches its maximum at full barrier transmissions. In contrast, CAR is
essentially a non-local effect which requires ``mixing'' of
trajectories for electrons going between two normal terminals with
those for electrons going deep into a superconductor and
describing the flow of Cooper pairs out of the contact area.
Provided there exists no normal electron reflection at both NS
interfaces such mixing does not occur, CAR vanishes and the only
remaining contribution to $G_{LR}$ in this case is one from DET.

For completeness, let us point out that the exact cancellation
between EC and CAR contributions \cite{Falci} can also be violated
by other means. One of them is simply to lift the spin degeneracy
in the system. This can be achieved, e.g., by considering NSN
structures with spin-active interfaces \cite{KZ07} or by using
ferromagnets (F) as normal metallic electrodes \cite{Yam,MF,Fazio}.
Experiments with FSF structures \cite{Beckmann} directly
demonstrated the dependence of the non-local conductance $G_{LR}$
on the polarization of F-electrodes.

Yet another way to avoid the cancellation between EC and CAR terms
already in the tunneling limit is to include interactions. This
idea has been put forward in Ref. \onlinecite{LY}. The effect of
electron-electron interactions on non-local conductance of NSN
devices -- in particular in the presence of disorder -- is an
interesting issue to be investigated further. Such
investigation is, however, beyond the scope of the present paper. Here we
only want to point out that interactions are not very likely to
play the dominant role in the experiments
\cite{Beckmann,Teun,Venkat}. This is because typical resistances
involved in these experiments were rather low and the
corresponding dimensionless conductances strongly exceeded unity.
Under such conditions the effect of Coulomb interactions on AR is
weak \cite{Z,HHK,GZ06} and a similar situation with EC and CAR can
be expected. In general, however, the combined effect of
electron-electron interactions and disorder on non-local
properties of NSN devices can be important.

In the absence of Coulomb interaction the effect of disorder on
non-local electron transport in NSN was recently considered in
Refs. \onlinecite{BG,Morten,Melin}. Brinkman and Golubov \cite{BG} employed the
quasiclassical formalism of Usadel equations and proceeded
perturbatively in the interface transmissions. They found that the
proximity effect in the normal electrodes in combination with
disorder can strongly enhance both EC and CAR contributions to the
non-local conductance. Duhot and Melin \cite{Melin} argued that
weak-localization-type of effects inside the superconductor may
influence non-local electron transport in NSN structures.
Morten {\it et al.} \cite{Morten}
considered a device with normal terminals attached to a
superconductor via an additional {\it normal} island (dot) and
analyzed this structure within the framework of the circuit
theory. In this paper we will extend and generalize this model 
by considering a {\it superconducting} dot attached to one
superconducting and two normal terminals as shown in Fig. 1.

Our main goal is to study the combined effect of proximity and
disorder {\it inside} the superconductor (dot). In addition, as it
was demonstrated in experiments \cite{Beckmann,Venkat},
non-equilibrium effects, such as charge imbalance, inside a superconducting 
electrode may play a significant role. These effects will be included into our
consideration too. We are going to show that the ``peaked'' temperature 
dependence of the non-local resistance $R_{LR}$ observed in the experiments 
\cite{Beckmann,Venkat} can be explained as a result of the competition
between charge imbalance and Andreev reflection. The crossover temperature
between these two processes $T^*$ (defined in Eq. (\ref{Tstar}) below) 
sets the position of the maximum in the dependence $R_{LR}(T)$.

In order to illustrate the main idea of our approach let us recall
that the exact cancellation between EC and CAR terms \cite{Falci}
occurs only at energies below the superconducting gap while at
higher energies (or in the normal state) CAR
vanishes and EC remains the only relevant mechanism of electron
transport.  It is clear, therefore, that including the proximity
effect due to the presence of normal electrodes immediately yields
non-zero subgap density of states inside the supercoducting
electrode which in turn should yield a decrease of CAR, thus eliminating
its compensation by EC and leaving the non-local conductance
$G_{LR}$ non-zero. This is precisely what we find. In order to
correctly account for the above effects it is necessary to proceed
non-perturbatively in tunneling and consider interface
conductances exceeding unity. Under these conditions the concept
of elastic cotunneling becomes irrelevant and it would be more
appropriate to speak about direct electron transfer between two
N-electrodes which includes processes of all orders in the
interface transmissions.

\begin{figure}
\includegraphics[width=7cm]{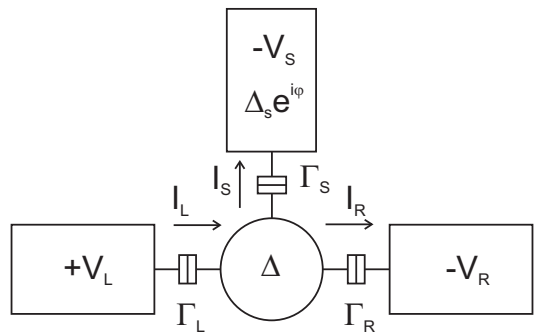}
\caption{Superconducting quantum dot coupled to two normal
and one superconducting leads.}
\end{figure}

The structure of our paper is as follows.  In Sec. 2 we will
introduce our model and outline the formalism to be used below. In
Sec. 3 we will evaluate the Green-Keldysh functions of our system
which will be used in Sec. 4 in order to derive the general
expressions for the non-local currents in our NSN device. In the limit of
low bias voltages these
general results are further analyzed in details in Sec. 5 where we also
illustrate their relation to previous theoretical works and to
experimental findings \cite{Beckmann,Venkat}.

\section{The model and basic formalism}

Below we will consider a chaotic superconducting quantum dot with
the mean level spacing $\delta$ connected to one superconducting
($S$) and two normal ($L$ and $R$) massive electrodes by means of
tunnel barriers. This structure is schematically shown in Fig. 1.
The typical size of the dot $d$ is supposed to be sufficiently
small, $d\lesssim \xi_0$, where $\xi_0$ is the superconducting
coherence length. Although we assume the channel transmissions of
all three junctions are small, $T_n^{(j)}\ll 1$ (here and below
$j=S,L,R$), their dimensionless conductances $g_j=2\sum_n
T_n^{(j)}$ can take any (large) value provided the number of
conducting channels is sufficiently large. The magnitudes of the
superconducting order parameters in the dot and the electrode $S$
are denoted respectively as $\Delta$ and $\Delta_S$. The phase
difference across the Josephson junction between the dot and the
$S$-electrode is denoted by $\varphi$.

Let us introduce the electron escape rate through the $j$-th
junction $\Gamma_j=g_j\delta/4\pi$. Below we will demonstrate that
transport properties of our system essentially depend on the
parameters $\Gamma_j/\Delta$. Our theoretical approach allows to
obtain the exact solution of our problem applicable for all values
of $\Gamma_j/\Delta$.

The Hamiltonian of our system reads
\begin{eqnarray}
H=H_{qd}+H_L+H_R+H_S+H_{TL}+H_{TR}+H_{TS},
\end{eqnarray}
where
\begin{eqnarray}
H_{qd}=\sum_{n,\;\alpha=\uparrow ,\downarrow } \xi_n \hat d_{\alpha n}^\dagger\hat d_{\alpha n}
+ \sum_n \big( \Delta\hat d_{\uparrow  n}^\dagger\hat d^\dagger_{\downarrow , -n}
+\,{\rm c.c.}\big)
\label{Hqd}
\end{eqnarray}
is the Hamiltonian of an isolated quantum dot,
\begin{equation}
H_{L,R}=\sum_{k,\alpha} \epsilon_{k\alpha} \hat c^\dagger_{k\alpha;L,R}\hat c_{k\alpha;L,R}
\end{equation}
represent the Hamiltonians of the left and right normal leads,
\begin{equation}
H_S=\sum_{k,\alpha} \epsilon_{k\alpha} \hat c^\dagger_{k\alpha;S}\hat c_{k\alpha;S} +\sum_k
\big(\Delta_S\hat c^\dagger_{\uparrow ,k,S}\hat
c^\dagger_{\downarrow ,-k,S} +\,{\rm c.c.}\big)
\end{equation}
is the Hamiltonian of the massive superconducting electrode and
\begin{equation}
H_{Tj}=\sum_{nk,\alpha} \big(t_{nk,j} \hat d^\dagger_{n\alpha}\hat c_{k\alpha;j} +
t^*_{nk,j}\hat c^\dagger_{k\alpha;j}\hat d_{n\alpha} \big)
\end{equation} define the
tunneling Hamiltonians for the junctions $j=S,L,R$.

Employing the Keldysh formalism we define non-equilibrium $4\times
4$ Green-Kedysh function of $n$-th energy level in the quantum
dot:
\begin{eqnarray}
\check G_n=\left(
\begin{array}{cc}
\hat G_n & \hat F_n \\
\hat F_n^\dagger & \hat G_n^\dagger
\end{array}
\right),
\end{eqnarray}
where
\begin{eqnarray}
\hat G_n=-i\left(
\begin{array}{cc}
\langle {\cal T}\hat d_{\uparrow n}(t_1) \hat d^\dagger_{\uparrow n}(t_2)\rangle &
-\langle \hat d^\dagger_{\uparrow n}(t_2)\hat d_{\uparrow n}(t_1)\rangle \\
\langle \hat d_{\uparrow n}(t_1) \hat d^\dagger_{\uparrow n}(t_2)\rangle &
\langle {\cal T}^{-1}\hat d_{\uparrow n}(t_1) \hat d^\dagger_{\uparrow n}(t_2)\rangle
\end{array}
\right), \nonumber
\end{eqnarray}
\begin{eqnarray}
\hat G^\dagger_n=-i\left(
\begin{array}{cc}
\langle {\cal T}\hat d^\dagger_{\downarrow n}(t_1) \hat d_{\downarrow n}(t_2)\rangle &
-\langle \hat d_{\downarrow n}(t_2)\hat d^\dagger_{\downarrow n}(t_1)\rangle  \\
\langle \hat d^\dagger_{\downarrow n}(t_1) \hat d_{\downarrow n}(t_2)\rangle &
\langle {\cal T}^{-1}\hat d^\dagger_{\downarrow n}(t_1) \hat d_{\downarrow n}(t_2)\rangle  \\
\end{array}
\right), \nonumber
\end{eqnarray}
\begin{eqnarray}
\hat F_n=-i\left(
\begin{array}{cc}
\langle {\cal T}\hat d_{\uparrow n}(t_1) \hat d_{\downarrow n}(t_2)\rangle &
-\langle \hat d_{\downarrow n}(t_2)\hat d_{\uparrow n}(t_1)\rangle \\
\langle \hat d_{\uparrow n}(t_1) \hat d_{\downarrow n}(t_2)\rangle &
\langle {\cal T}^{-1}\hat d_{\uparrow n}(t_1) \hat d_{\downarrow n}(t_2)\rangle
\end{array}
\right), \nonumber
\end{eqnarray}
\begin{eqnarray}
\hat F^\dagger_n = -i\left(
\begin{array}{cc}
\langle {\cal T}\hat d^\dagger_{\downarrow n}(t_1) \hat d^\dagger_{\uparrow n}(t_2)\rangle &
-\langle  \hat d^\dagger_{\uparrow n}(t_2)\hat d^\dagger_{\downarrow n}(t_1)\rangle \\
\langle \hat d^\dagger_{\downarrow n}(t_1) \hat d^\dagger_{\uparrow n}(t_2)\rangle &
\langle {\cal T}^{-1}\hat d^\dagger_{\downarrow n}(t_1) \hat d^\dagger_{\uparrow n}(t_2)\rangle
\end{array}
\right). \nonumber
\end{eqnarray}
Note that here we do not introduce the off-diagonal elements of
the Green function $\check G_{nm}$. In the next section we will
demonstrate that these off-diagonal elements vanish, $\check
G_{nm}=0$ for $n\neq m$, provided the dot is fully chaotic and all
channel transmissions are small, $T_n^{(j)}\ll 1$. The Green
functions of the leads are defined analogously.

The current through the left tunnel junction is expressed as
\begin{eqnarray}
I_L&=&\frac{e}{2} \sum_{nk} |t_{nk}|^2\int\frac{dE}{2\pi}
\,{\rm tr}\, \big[ \check G_n(E)\check\Lambda\check G_{k,L}(E)
\nonumber\\ &&
-\, \check G_{k,L}(E)\check \Lambda \check G_n(E) \big].
\label{IL11}
\end{eqnarray}
Here $\Lambda$ is the $4\times 4$ diagonal matrix with the matrix
elements $\Lambda_{11}=-1$, $\Lambda_{22}=1$, $\Lambda_{33}=1$ and
$\Lambda_{44}=-1$. The functions $\check G_n(E)$ and $\check
G_L(E)$ are the Fourier components of the Green-Keldysh functions
for the dot and the left lead respectively. The currents across
the right junction and across the Josephson junction between the dot and
the superconducting electrode are defined analogously.

In our subsequent calculation we will make use of the fact that
coupling of the $n$-th energy level of a chaotic dot to the leads
does not depend on the number $n$, i.e. it remains the same for
all levels. Hence, the effective level width $\delta\epsilon =
\Gamma_L+\Gamma_R+\Gamma_S$ is also the same for all the dot
levels. This observation enables us to first evaluate the $4\times
4$ Green-Keldysh functions for each single energy level, then
calculate its contribution to the current and afterwards perform a
summation over all energy levels. This program will be
accomplished below.

\section{Green-Keldysh functions}

The Green-Keldysh functions of the dot $\check G_{ln}$ obey the
Dyson equation
\begin{eqnarray}
\sum_l\big[ \check G_{n,qd}^{-1}\delta_{ml}-\check \Sigma_L^{ml}-\check \Sigma_R^{ml}-\check\Sigma_S^{ml} \big]
\check G_{ln}=\check 1\delta_{mn},
\end{eqnarray}
where
\begin{eqnarray}
\check G_{n,qd}^{-1}=\left(
\begin{array}{cccc}
E-\xi_n & 0 & -\Delta & 0 \\
0 & -E+\xi_n & 0 & \Delta \\
-\Delta & 0 & E+\xi_n & 0 \\
0 & \Delta & 0 & -E-\xi_n
\end{array}
\right),
\end{eqnarray}
and
\begin{eqnarray}
\check\Sigma^{mn}_j= \sum_k t^*_{km,j}t_{kn,j}\check\Lambda\check G_{k,j}(E)\check\Lambda.
\end{eqnarray}
is the self-energy of the $j-$th junction.

In a chaotic quantum dot off-diagonal matrix elements of any
operator between the $m-$th and the $n-$th energy levels tend to
zero provided their energies are not too far from each other,
$|\xi_n-\xi_m|\lesssim D/d^2$, where $D$ is the diffusion
coefficient for electrons inside the dot\cite{AGB}. In addition, under these
conditions the diagonal matrix elements do not depend on the level
number $n$, i.e. $\langle n|\hat A|n\rangle={\rm const}$\cite{AGB}. Hence,
we obtain
\begin{eqnarray}
\check\Sigma^{mn}_j=\delta_{mn} \check\Sigma_j,
\end{eqnarray}
where
\begin{eqnarray}
\check\Sigma_j=|t_j|^2 \check\Lambda\sum_k\check G_{k,j}(E)\check\Lambda.
\label{Sigma1}
\end{eqnarray}
Then the Dyson equation for the Green-Keldysh function acquires
the form
\begin{eqnarray}
\big[ \check G_{n,qd}^{-1}-\check \Sigma_L-\check \Sigma_R-\check\Sigma_S \big]
\check G_{n}=\check 1.
\label{Dyson}
\end{eqnarray}

Performing the summation over $k$ in Eq. (\ref{Sigma1}) with the
known expressions for the Green-Keldysh functions of the leads is
straightforward. As a result, the self-energies for the junctions
between the dot and the normal leads take the form
\begin{eqnarray}
\check\Sigma_{L,R}=\frac{\Gamma_{L,R}}{2i}\left(
\begin{array}{cc}
\hat\sigma_z\hat Q(E\pm eV_{L,R}) & 0 \\
0 & \hat\sigma_z\hat Q(E\mp eV_{L,R})
\end{array}
\right)
\label{slr}
\end{eqnarray}
where $\hat\sigma_z$ is the Pauli matrix and $\hat Q$ is the $2\times
2$ matrix which reads
\begin{eqnarray}
\hat Q(E)=\left(
\begin{array}{cc}
1-2n(E) & 2n(E) \\
2-2n(E) & -1+2n(E)
\end{array}
\right).
\end{eqnarray}
Here $n(E)=1/(1+e^{E/T})$ is the Fermi function.

The self-energy $\check \Sigma_S$ for the Josephson junction
between the dot and the $S$-electrode, though somewhat more
involved, is evaluated analogously. Combining the resulting
expression for $\check \Sigma_S$ with Eq. (\ref{slr}) we obtain
\begin{eqnarray}
&& \check \Sigma = \check \Sigma_L+\check \Sigma_R+\check \Sigma_S
\nonumber\\ &&
=\,-\frac{\Gamma_SF_S(E)}{2}\left(
\begin{array}{cc}
E/\Delta_S &  -e^{i\varphi} \\
-e^{-i\varphi} &  E/\Delta_S
\end{array}
\right)\otimes \hat\sigma_z 
\nonumber\\ &&
-\,i\frac{\Gamma(E)}{2}
\left(\begin{array}{cc} 1 & 0 \\ 0 &
0 \end{array}\right)\otimes \hat\sigma_z\hat Q_{qd}(E)
\nonumber\\ &&
-\,i\frac{\Gamma(E)}{2}
\left(\begin{array}{cc} 0 & 0 \\ 0 &
1 \end{array}\right)\otimes
\hat\sigma_z\hat Q'_{qd}(E)
\nonumber\\ &&
+\,i\frac{\Gamma_S}{2}\frac{\Delta_SN_S(E)}{E} \left(
\begin{array}{cc} 0 & e^{i\varphi} \\ e^{-i\varphi} & 0
\end{array} \right)\otimes \hat\sigma_z\hat Q_S(E),
\label{stot}
\end{eqnarray}
where we denoted
$F_S(E)=\frac{\Delta_S\theta(\Delta_S-|E|)}{\sqrt{\Delta_S^2-E^2}}$,
introduced the density of states in the superconductor
$N_S(E)=\frac{|E|\theta(|E|-\Delta_S)}{\sqrt{E^2-\Delta^2}}$ and
defined
\begin{eqnarray}
\Gamma(E)=\Gamma_L+\Gamma_R+\Gamma_SN_S(E),
\end{eqnarray}
as the total escape rate of an electron from the dot through all
three barriers. The $\hat Q$-matrices in Eq. (\ref{stot}) read
\begin{eqnarray}
\hat Q_{qd}&=&\left(
\begin{array}{cc} 1-2n_{qd} & 2n_{qd} \\ 2-2n_{qd} & -1+2n_{qd}  \end{array}
\right),
\label{Qqd}
\end{eqnarray}
where
\begin{eqnarray}
&& n_{qd}(E)=\frac{\Gamma_L}{\Gamma(E)}n(E+eV_L)+\frac{\Gamma_R}{\Gamma(E)}n(E-eV_R)
\nonumber\\ &&
+\, \frac{\Gamma_S}{2\Gamma(E)}\big(N_S(E)+\theta(|E|-\Delta_S)\big)n(E-eV_S)
\nonumber\\ &&
+\, \frac{\Gamma_S}{2\Gamma(E)}\big(N_S(E)-\theta(|E|-\Delta_S)\big)n(E+eV_S)
\label{nqd}
\end{eqnarray}
is the distribution function in the quantum dot,
\begin{equation}
\hat Q'_{qd}(E,V_L,V_R,V_S)=\hat Q_{qd}(E,-V_L,-V_R,-V_S)
\end{equation}
and
\begin{equation} \hat Q_S(E)=\big[\hat Q(E-eV_S)+\hat
Q(E+eV_S)\big]/2.
\end{equation}

We are now in a position to evaluate the Green-Keldysh function
with the aid of the Dyson equation (\ref{Dyson}). The derivation
is facilitated by the normalization condition for the $\hat
Q$-matrices, $\hat Q^2=1$, as well as by the property
$$\hat Q_i\hat Q_j=\hat 1-\hat Q_i+\hat Q_j.
$$
After some algebra we finally arrive at the following expression
for the dot Green-Keldysh function:
\begin{eqnarray}
&& \check G_n(E,\xi_n)= \frac{1}{2}\left(\tilde G_{n}^R+\tilde G^A_n\right)\otimes \hat \sigma_z
\nonumber\\ &&
+\,i\frac{\Gamma_SN_S(E)\Delta_S}{2E} \tilde G_{n}^R
\left(\begin{array}{cc} 0 & e^{i\varphi} \\ e^{-i\varphi} & 0  \end{array}\right)
\tilde G_{n}^A\otimes \hat Q_{S}(E)\hat\sigma_z
\nonumber\\ &&
-\,i\frac{\Gamma(E)}{2} \tilde G_{n}^R
\left(\begin{array}{cc} 1 & 0 \\ 0 & 0  \end{array}\right)
\tilde G_{n}^A\otimes \hat Q_{qd}(E)\hat\sigma_z
\nonumber\\ &&
-\,i\frac{\Gamma(E)}{2} \tilde G_{n}^R
\left(\begin{array}{cc} 0 & 0 \\ 0 & 1  \end{array}\right)
\tilde G_{n}^A\otimes \hat Q'_{qd}(E)\hat\sigma_z.
\label{G}
\end{eqnarray}
Here we defined the $2\times 2$ matrix retarded and advanced Green functions
\begin{eqnarray}
\tilde G_n^{R,A}(E,\xi_n,\varphi)=\left(
\begin{array}{cc}
G_{R,A}(E,\xi_n,\varphi) & F_{R,A}(E,\xi_n,\varphi) \\ F_{R,A}(E,\xi_n,-\varphi) & G_{R,A}(E,-\xi_n,\varphi)
\end{array}
\right),
\nonumber
\end{eqnarray}
where
\begin{eqnarray}
G_{R,A}(E,\xi,\varphi)&=&
\frac{E+\xi+\frac{\Gamma_SF_S(E) E}{2\Delta_S}\pm i\frac{\Gamma(E)}{2}}{P_{R,A}(E,\xi,\varphi)},
\label{GF0}\\
F_{R,A}(E,\xi,\varphi)&=& \frac{\Delta
+\frac{\Gamma_Se^{i\varphi}}{2}\left(F_S(E)  \pm
i\frac{N_S(E)\Delta_S}{E}\right)} {P_{R,A}(E,\xi,\varphi)}
\label{GF}
\end{eqnarray}
are respectively the normal and anomalous retarded and advanced
Green functions of the superconducting quantum dot and
\begin{eqnarray}
&& P_{R,A}(E,\xi,\varphi)=\left( E+\frac{\Gamma_SF_S(E) E}{2\Delta_S}\pm i\frac{\Gamma(E)}{2} \right)^2
\nonumber\\ &&
-\,\xi^2-\left( \Delta +\frac{\Gamma_Se^{i\varphi}}{2}\left(F_S(E)  \pm i\frac{N_S(E)\Delta_S}{E}\right) \right)
\nonumber\\ &&\times
\left( \Delta +\frac{\Gamma_Se^{-i\varphi}}{2}\left(F_S(E)  \pm
    i\frac{N_S(E)\Delta_S}{E}\right) \right).
\label{PRA}
\end{eqnarray}
One can easily verify that the retarded and advanced Green
functions are linked to each other by the standard relations
$G_A(E,\xi,\varphi)=G_R^*(E,\xi,\varphi)$ and
$F_A(E,\xi,\varphi)=F_R^*(E,\xi,-\varphi)$.

For completeness, we also present the self-consistency equation
which controls the magnitude of the order parameter in the dot:
\begin{eqnarray}
\Delta &=& -\frac{\lambda\Gamma_S}{2\delta}\sum_n\int\frac{dE}{2\pi}
F_S(E)\,{\rm sign}\,E\;
\nonumber\\ &&\times\,
\big[1-n(E+eV_S)-n(E-eV_S)\big]
\nonumber\\ &&\times\,
\big[ \cos 2\varphi\;G_R(E,\xi_n,\varphi)G_R^*(E,-\xi_n,\varphi) 
\nonumber\\ &&
+\, F_R(E,\xi_n,\varphi)F_R^*(E,\xi_n,-\varphi) \big]
\nonumber\\ &&
+\,\frac{\lambda}{\delta}\sum_n\int\frac{dE}{2\pi}
\Gamma(E)
\,{\rm Re}\,\big[e^{i\varphi}G_R(E,\xi_n,\varphi)
\nonumber\\ &&\times\,
F_R^*(E,\xi_n,-\varphi)\big]\big[1-2n_{qd}(E)\big].
\label{selfcons}
\end{eqnarray}
where $\lambda$ is the BCS coupling constant. 
In general, the superconducting order parameter inside the dot should be 
determined self-consistently with the aid of Eq. (\ref{selfcons}).
Here we avoid this complication and set $\Delta$ equal to a constant.
This assumption is justified if, for instance, the coupling between 
the dot and the superconducting lead is much stronger 
than that between the dot and the normal
leads, $\Gamma_S\gg \Gamma_L,\Gamma_R$. If, in addition, we assume
that both the dot and the superconducting lead are made of the same material,
it would be appropriate to set $\Delta=\Delta_S$ at all temperatures and sufficiently
low bias voltages.

\section{Non-local currents}

We now make use of the above general results and evaluate the
currents across both NS interfaces of our device. Combining Eqs.
(\ref{IL11}) and (\ref{Sigma1}) we express the current across the
left interface in the form
\begin{eqnarray}
I_L&=& \frac{e}{2}\sum_n\int\frac{dE}{2\pi}\,{\rm tr}\,
\left(\big[\check\Sigma_L(E),\check\Lambda\big]\check G_n(E)\right).
\end{eqnarray}
An analogous formula is obtained for the current in the right
junction $I_R$. Substituting the results for the Green functions
and self-energies derived in the previous section into the above
expressions for the currents and setting $V_S=0$, we
obtain
\begin{eqnarray}
I_L&=& I_{LS}(V_L)+\left(
2\frac{\Gamma_L}{\Gamma_R}+1\right)I_{CAR}(V_L)+I_{DET}(V_L) \nonumber\\
&& +\, I_{DET}(V_R) - I_{CAR}(V_R),
\label{IL21}\\
I_R&=& I_{RS}(V_R)+\left(
2\frac{\Gamma_R}{\Gamma_L}+1\right)I_{CAR}(V_R) + I_{DET}(V_R) \nonumber\\
&& +\,I_{DET}(V_L)-I_{CAR}(V_L), \label{nonlcurr}
\end{eqnarray}
where we defined
\begin{eqnarray}
&& I_{LS}(V) = \frac{e\Gamma_L\Gamma_S}{\pi}\sum_n \int dE N_S(E)
\nonumber\\ &&\times\,
\bigg\{  |G_R(E,\xi_n,\varphi)|^2 + |F_R(E,\xi_n,\varphi)|^2
\nonumber\\ &&
-\,\frac{2\Delta_S}{E}\,{\rm Re}\,\big[ G_R^*(E,\xi_n,\varphi)F_R(E,\xi_n,-\varphi)e^{i\varphi} \big]\bigg\}
\nonumber\\ &&\times\,
\left[n(E- eV)-n(E)\right],
\end{eqnarray}
\begin{eqnarray}
&& I_{RS}(V) = \frac{e\Gamma_R\Gamma_S}{\pi}\sum_n \int dE N_S(E)
\nonumber\\ &&\times\,
\bigg\{  |G_R(E,\xi_n,\varphi)|^2 + |F_R(E,\xi_n,\varphi)|^2
\nonumber\\ &&
-\,\frac{2\Delta_S}{E}\,{\rm Re}\,\big[ G_R(E,\xi_n,\varphi)F_R^*(E,\xi_n,\varphi)e^{i\varphi} \big]\bigg\}
\nonumber\\ &&\times\,
\left[n(E- eV)-n(E)\right],
\end{eqnarray}
\begin{eqnarray}
I_{DET}(V)&=&\frac{e\Gamma_L\Gamma_R}{\pi}\sum_n\int dE
|G_R(E,\xi_n)|^2 \nonumber\\ &&\times\, \big[ n(E-eV)-n(E) \big],
\label{DETcurr}\\
I_{CAR}(V)&=&\frac{e\Gamma_L\Gamma_R}{\pi}\sum_n\int dE
|F_R(E,\xi_n)|^2 \nonumber\\ &&\times\, \big[ n(E-eV)-n(E) \big].
\label{CARcurr}
\end{eqnarray}
Eqs. (\ref{nonlcurr})-(\ref{CARcurr}) fully determine the currents
across the left and the right NS interfaces and represent the
central result of our paper.

The current $I_{DET}(V)$ accounts for direct electron transfer
between two normal terminals. This current differs from zero also
in the normal state of our system. In contrast, $I_{CAR}(V)$
describes the contribution from crossed Andreev reflection which
vanishes in the normal limit. The contributions $I_{LS}$ and
$I_{RS}$ contain terms which can be interpreted in a 
similar, though slightly more complicated manner since they
originate from the Josephson junction between the superconductors
and not from the NS interface. If, just for illustration, we put
$\Delta_S=0$ we immediately get
$I_{LS}=(\Gamma_S/\Gamma_R)(I_{DET}+I_{CAR})$ and
$I_{RS}=(\Gamma_S/\Gamma_L)(I_{DET}+I_{CAR})$.

We note that the possibility to decompose the currents $I_{L,R}$
into the sum of partial currents (\ref{nonlcurr}), each of which depending only
on either $V_L$ or $V_R$ (but not on both) 
is due to the fact that the distribution function in the
quantum dot $n_{qd}(E)$ (\ref{nqd}) is represented as a linear
combination of the distribution functions of the leads. This
feature is similar to that of ballistic NSN devices
\cite{KZ06,KZ07}.

\subsection{Normal state}

Let us analyze the above general expressions for the current.
Considering first the trivial limit of a normal system
$\Delta=\Delta_S=0$ we obtain
\begin{eqnarray}
n_{qd}=\frac{\Gamma_Ln(E+eV_L)+\Gamma_Rn(E-eV_R)+\Gamma_Sn(E)}{\Gamma_L+\Gamma_R+\Gamma_S}.
\label{nqd1}
\end{eqnarray}
The current through the left junction is defined by a simple formula
\begin{eqnarray}
I_L=\frac{1}{eR_L}\int dE\,\big[n_{qd}(E)-n(E+eV_L)\big].
\end{eqnarray}
The expression for $I_R$ is similar. Evaluating the integral over $E$
we obtain
\begin{eqnarray}
\left(\begin{array}{c} I_L \\ I_R \end{array}\right) =
\left(\begin{array}{cc} G_{LL}^N & G_{LR}^N \\ G_{RL}^N & G_{RR}^N  \end{array}\right)
\left(\begin{array}{c} V_L \\ V_R \end{array}\right)
\label{cur}
\end{eqnarray}
where
\begin{eqnarray}
&& G_{LL(RR)}^N=\frac{R_S+R_{R(L)}}{R_LR_R+R_LR_S+R_RR_S}, \label{GLLN}
\\ &&
G_{LR}^{N}=G_{RL}^N = \frac{R_S}{R_LR_S+R_RR_S+R_LR_R}. \label{GN}
\end{eqnarray}
In this limit both local and non-local differential conductances
remain voltage-independent.

In the experiments one often measures the non-local resistance 
\begin{eqnarray}
R_{LR}=-\left.\frac{\partial V_R}{\partial I_L}\right|_{I_R=0}=\frac{G_{LR}}{G_{LL}G_{RR}-G_{LR}^2}.
\label{Rnl}
\end{eqnarray}
From Eqs. (\ref{cur}) we obtain 
\begin{eqnarray}
R_{LR}^N=R_S.
\end{eqnarray}

\subsection{Charge imbalance}

Charge imbalance\cite{CT} is a non-equilibrium phenomenon which is known to
cause a number of interesting non-local effects in superconductors. This
phenomenon is also of importance in connection  with non-local
electron transport in NSN hybrid structures discussed here. 
In particular, it was argued \cite{Beckmann,Venkat} that charge imbalance
might be responsible for certain features of the non-local conductance
observed in experiments. Our approach
allows to fully account for this phenomenon and its impact
on non-local transport in the system under consideration. In this subsection
we briefly illustrate the key physics associated with charge imbalance 
in our system. 

Just for the sake of illustration let us for a moment set $\Delta_S=0$ 
and assume $\Gamma_L,\Gamma_R,\Gamma_S\ll \Delta$. 
In this regime the current $I_L$ (\ref{IL21}) across the left junction takes the form\cite{Tinkham}
\begin{eqnarray}
 I_L&=&\frac{1}{eR_L}\int dE\,\frac{|E|\theta(|E|-\Delta)}{\sqrt{E^2-\Delta^2}}\big[n(E)-n(E+eV_L)\big]
\nonumber\\ &&
+\,\frac{1}{eR_L}\int dE\,\theta(|E|-\Delta)\,\tilde n_{qd}(E),
\label{IL1}
\end{eqnarray}
where $\tilde n_{qd}(E)$
is the asymmetric part of the distribution function responsible for 
charge imbalance \cite{Tinkham}. 
For the system under consideration $\tilde n_{qd}(E)$ reads \cite{VG}
\begin{eqnarray}
\tilde n_{qd}(E)=\frac{\sqrt{E^2-\Delta^2}}{|E|}\frac{n_{qd}(E)+n_{qd}(-E)-1}{2},
\end{eqnarray}
and $n_{qd}(E)$ is given by Eq. (\ref{nqd1}). Then for local and non-local zero
bias conductances one obtains 
\begin{eqnarray}
 G_{LL(RR)} = \frac{1}{R_{L(R)}}\int dE\frac{\theta(|E|-\Delta)}{4T\cosh^2\frac{E}{2T}}
\left(\frac{|E|}{\sqrt{E^2-\Delta^2}}
\right.
\nonumber\\ 
\left.
-\,\frac{R_{R(L)}R_S}{R_LR_R+R_LR_S+R_RR_S}\frac{\sqrt{E^2-\Delta^2}}{|E|}\right),
\label{GLLapp}
\end{eqnarray}
\begin{eqnarray}
 G_{LR}= G_{LR}^N\int dE\frac{\theta(|E|-\Delta)}{4T\cosh^2\frac{E}{2T}}
\frac{\sqrt{E^2-\Delta^2}}{|E|}.
\label{GLRapp}
\end{eqnarray}
Note that in this regime the non-local conductance is solely due to charge
imbalance being fully determined by the second term in Eq. (\ref{IL1}).
At low temperatures we have $G_{LL},G_{RR},G_{LR}\propto e^{-\Delta/T}$.
Hence, in the situation considered in this subsection at $T\to 0$
the non-local resistance (\ref{Rnl}) should diverge as $R_{LR}\propto R_Se^{\Delta/T}$.

\subsection{General case}

\begin{figure}
\begin{tabular}{ll}
\includegraphics[width=4cm]{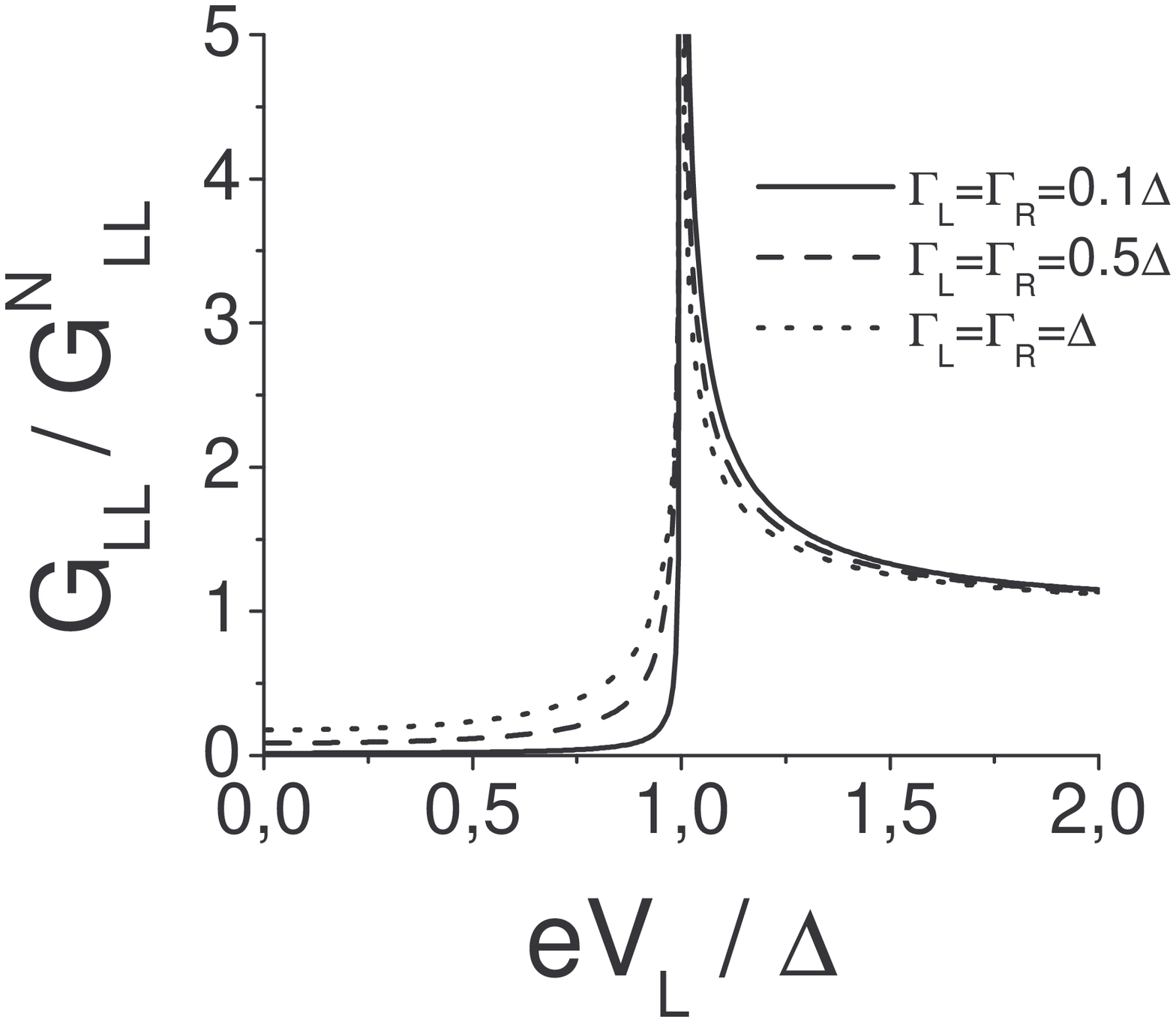}\hspace{0.3cm} &
\includegraphics[width=4cm]{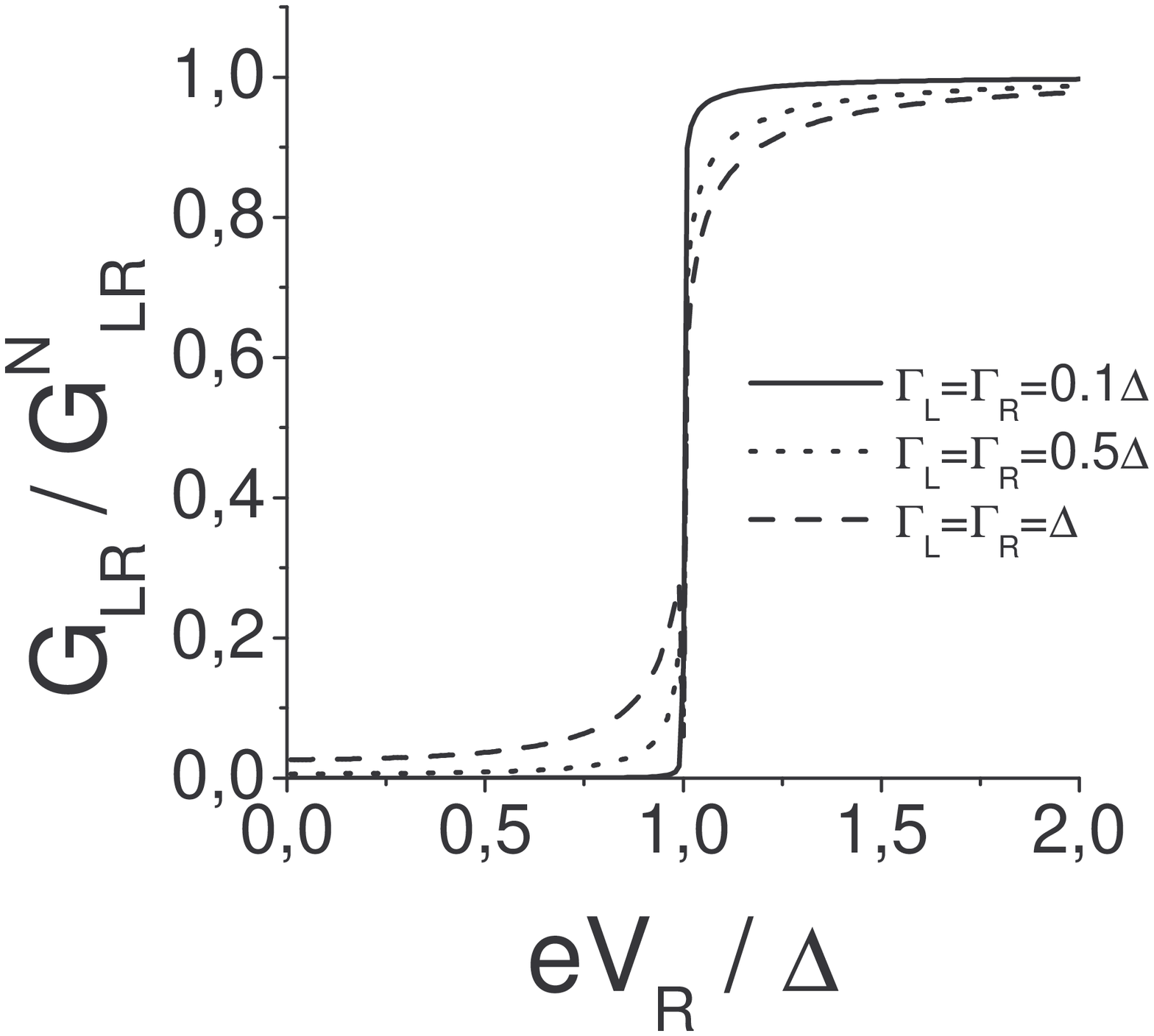}
\end{tabular}
\caption{Normalized local $G_{LL}(V_R)/G^N_{LL}$ and non-local $G_{LR}(V_R)/G^N_{LR}$
differential conductances
as a function of applied voltage $V_R$ at low
temperature $T\ll\Delta$ and at different tunneling rates
$\Gamma_L=\Gamma_R$. Here we set $\Delta_S=\Delta$ and
$\Gamma_S=10\Delta$.} \label{FigGamma}
\end{figure}

Now let us return to the case $\Delta_S \neq 0$. 
With the aid of Eqs. 
(\ref{DETcurr}), (\ref{CARcurr}) we determine the
differential conductance $G_{LR}$ which is presented in Fig.
\ref{FigGamma} as a function of applied voltage $V_R$ for $\Delta=\Delta_S$
and 
different values of the tunneling rates
$\Gamma_L=\Gamma_R$ as compared to $\Delta$. We observe that at
subgap voltages $eV_R<\Delta$ the magnitude of the normalized
non-local conductance $G_{LR}/G_N$ increases with increasing
$(\Gamma_L+\Gamma_R)/\Delta$. Such dependence is quite natural
because exactly the same ratio controls the strength of the
proximity effect in our system. As we have already discussed, with
increasing value of the ratio $(\Gamma_L+\Gamma_R)/\Delta$ the
proximity-induced subgap electron density of states increases, the
difference between DET and CAR contributions grows and, hence,
$G_{LR}$ becomes bigger.

\begin{figure}
\begin{tabular}{cc}
\includegraphics[width=4cm]{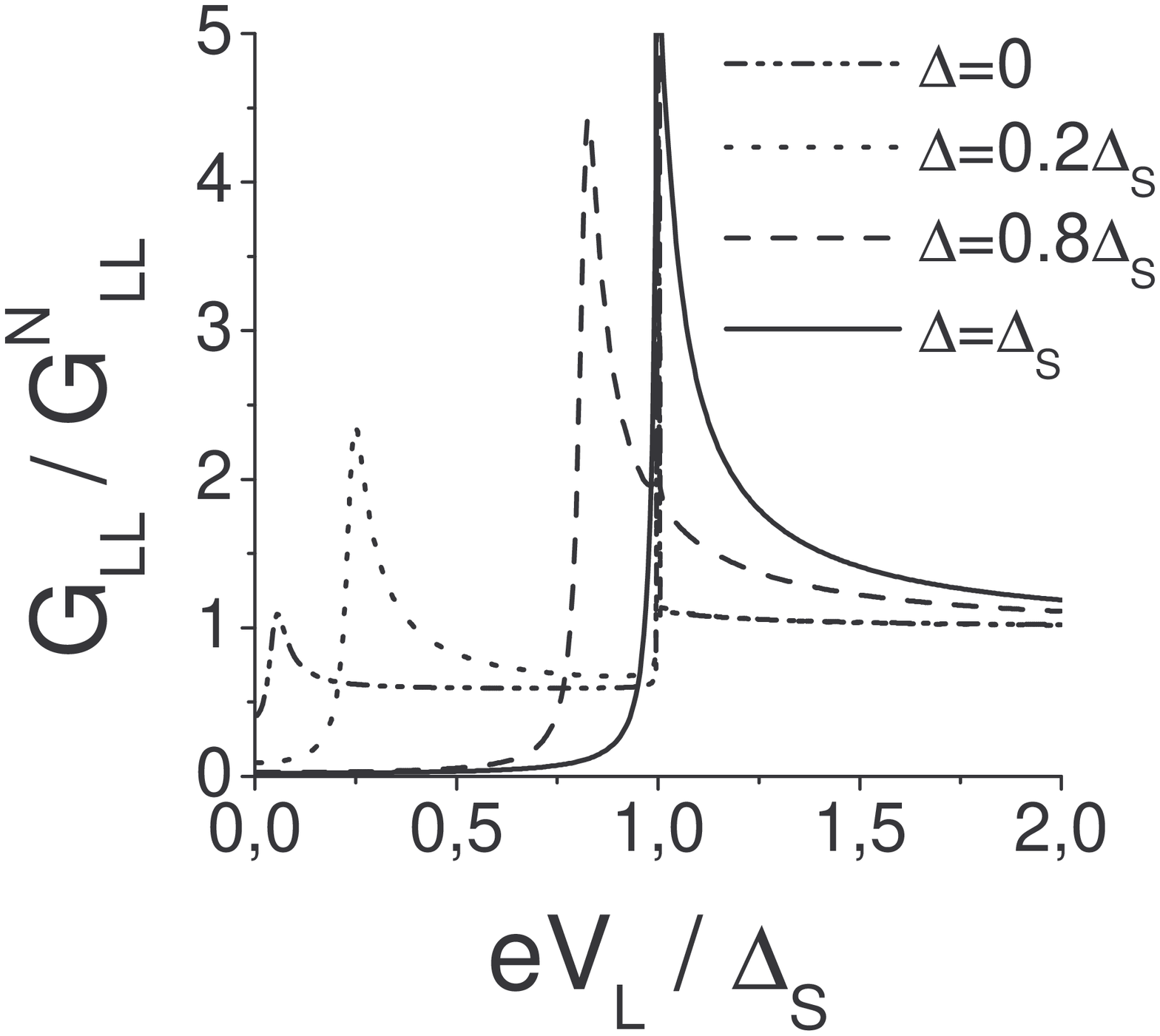}
\hspace{0.3cm}
&
\includegraphics[width=4cm]{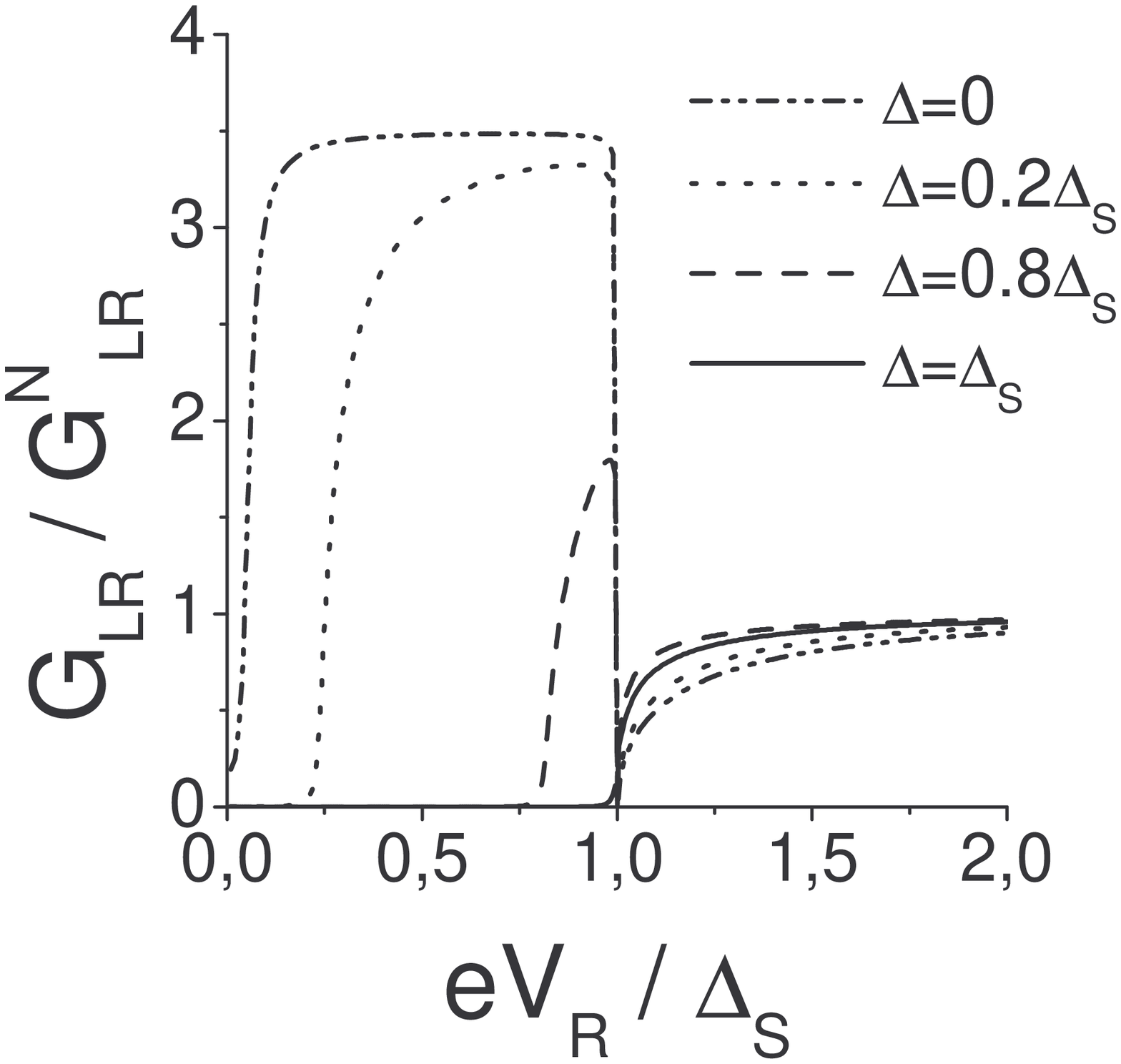}
\end{tabular}
\caption{The same as in Fig. \ref{FigGamma} for different values
of the dot gap $\Delta$. Here we set
$\Gamma_L=\Gamma_R=0.02\Delta_S$ and $\Gamma_S=0.1\Delta_S$.}
\label{FigDelta}
\end{figure}

Fig. \ref{FigDelta} illustrates the effect of the dot order parameter
$\Delta$ on the non-local conductance. At high voltages 
$eV_R\gtrsim \Delta_S$ we recover the normal state value (\ref{GN}) 
while at intermediate values $\Delta \lesssim eV_R
\lesssim \Delta_S$ we find $G_{LR}\approx 1/(R_L+R_R)$.
For $eV_R<\Delta$ the
conductance $G_{LR}$ progressively increases with decreasing ratio
$\Delta /\Delta_S$ and eventually reaches the maximum in the limit
$\Delta =0$ in which case the results \cite{Morten} are
reproduced. 

In order to demonstrate an important difference in the
low voltage behavior of $G_{LR}$ for superconducting and normal quantum dots
in Fig. \ref{FigDelta} we deliberately chose small values of tunneling rates 
$\Gamma_{L,R} \ll \Delta$. We observe that in the superconducting case  
$G_{LR}$ essentially vanishes 
at $eV_R < \Delta$ in accordance with Ref. \onlinecite{Falci}, 
while for normal quantum dots \cite{Morten} $G_{LR}$ remains non-zero
even at $V_R \to 0$.

\section{Zero-bias conductances}

Let us now consider the behavior of the conductance
matrix in the limit of low voltages in more details. In the zero
bias regime currents flowing through the system are low, i.e. we
can set $\varphi=0$. Similarly to Ref. \onlinecite{Falci} at low
voltages the expression for current $I_L$ can be split into three
different contributions
\begin{eqnarray}
I_L= G_{A}V_L + G_{DET}(V_L+V_R)+G_{CAR}(V_L-V_R).
\label{cnds}
\end{eqnarray}
Here $G_{A}$ is (local) Andreev conductance of the left NS
barrier, $G_{DET}$ and
$G_{CAR}$ are respectively DET and CAR contributions to the zero
bias conductance matrix. 

\subsection{Zero temperature limit}

In the limit of zero temperature $T \to
0$ from Eqs. (\ref{nonlcurr})-(\ref{CARcurr}) we obtain
\begin{eqnarray}
G_{A}= \frac{e^2}{\pi}\sum_n
\frac{2\Gamma_L^2\left(\Delta^2+\Gamma_S\Delta +
\frac{\Gamma_S^2}{2}\right)} {\left(
\xi_n^2+\Delta^2+\frac{(\Gamma_L+\Gamma_R)^2+\Gamma_S^2}{4}+\Gamma_S\Delta
\right)^2}, \label{GDA1}
\end{eqnarray}
\begin{eqnarray}
G_{DET}= \frac{e^2}{\pi}\sum_n
\frac{\Gamma_L\Gamma_R\left(\xi_n^2+
\frac{(\Gamma_L+\Gamma_R)^2}{4}\right)} {\left(
\xi_n^2+\Delta^2+\frac{(\Gamma_L+\Gamma_R)^2+\Gamma_S^2}{4}+\Gamma_S\Delta
\right)^2}, \label{GEC1}
\end{eqnarray}
\begin{eqnarray}
G_{CAR}= \frac{e^2}{\pi}\sum_n
\frac{\Gamma_L\Gamma_R\left(\Delta^2+\Gamma_S\Delta +
\frac{\Gamma_S^2}{4}\right)} {\left(
\xi_n^2+\Delta^2+\frac{(\Gamma_L+\Gamma_R)^2+\Gamma_S^2}{4}+\Gamma_S\Delta
\right)^2}. \label{GCA1}
\end{eqnarray}

Let us analyze the above expressions in different physical limits.
We first put $\Gamma_S=0$ and $\Delta=0$, i.e. we consider a
normal quantum dot isolated from the superconducting electrode.
Then we obviously find $G_{A}=G_{CAR}=0$, while for $G_{DET}$ we
obtain
\begin{eqnarray}
G_{DET}=\frac{e^2}{\pi}\sum_n \frac{\Gamma_L\Gamma_R}{
\xi_n^2+\frac{(\Gamma_L+\Gamma_R)^2}{4} }.
\end{eqnarray}
Comparing this expression to the Landauer formula we immediately
conclude that each energy level of the dot effectively corresponds
to one conducting channel with transmission
\begin{eqnarray}
\tau_n=\frac{\Gamma_L\Gamma_R}{\xi_n^2 +\frac{(\Gamma_L+\Gamma_R)^2}{4}}.
\label{tau1}
\end{eqnarray}
Considering a big metallic quantum dot we can replace  the sum
over energy states by the integral $\sum_n \to
\frac{1}{\delta}\int d\xi$. Making use of the relation between the
tunneling rates and the the junction resistances,
$\Gamma_{L,R}=\delta /2e^2R_{L,R}$, we reproduce the standard
result
\begin{eqnarray}
G_{DET}=\frac{1}{R_L+R_R},
\end{eqnarray}
i.e. in this case DET contribution simply reduces to the Ohm's
law.

Next we put $\Gamma_R=0$ and consider a superconducting dot
coupled to one normal and one superconducting lead. In this case
one trivially gets $G_{DET}=G_{CAR}=0$. Provided
$\Gamma_L,\Gamma_S\gg \Delta$ the dot can be viewed as a
point-like scatterer with the following set of transmission
probabilities (cf. Eq. (\ref{tau1}))
\begin{eqnarray}
\tilde\tau_n=\frac{\Gamma_L\Gamma_S}{\xi_n^2 +\frac{(\Gamma_L+\Gamma_S)^2}{4}}.
\end{eqnarray}
We note that, although the channel transmissions of NS interfaces
remain small, effective transmissions $\tilde\tau_n$ are not
necessarily small. The Andreev conductance in this limit becomes
\begin{eqnarray}
G_{A}= \frac{e^2}{2\pi}\sum_n
\frac{\Gamma_L^2\Gamma_S^2}{\left( \xi_n^2+\frac{\Gamma_L^2+\Gamma_S^2}{4}\right)^2}.
\end{eqnarray}
One can verify that this expression can be cast to the familiar
form \cite{BTK,B}
\begin{eqnarray}
G_{A}=\frac{e^2}{\pi}\sum_n\frac{2\tilde\tau_n^2}{(2-\tilde\tau_n)^2}.
\end{eqnarray}

In a general case of metallic quantum dots one can perform the
summation over $\xi_n$ in Eqs. (\ref{GDA1}-\ref{GCA1}) and arrive
at the following explicit expressions
\begin{eqnarray}
G_{A}= \frac{e^2}{4\pi}g_L^2 \frac{{\cal B}}{{\cal K}^{3/2}},
\label{GDA}
\end{eqnarray}
\begin{eqnarray}
G_{DET}= \frac{e^2}{8\pi} g_Lg_R \frac{{\cal
B}+(g_L+g_R)^2/2}{{\cal K}^{3/2}}, \label{GEC}
\end{eqnarray}
\begin{eqnarray}
G_{CAR}= \frac{e^2}{8\pi}g_Lg_R \frac{{\cal B}}{{\cal K}^{3/2}}.
\label{GCA}
\end{eqnarray}
where
\begin{eqnarray}
{\cal
B}=\frac{16\pi^2\Delta^2}{\delta^2}+g_S\frac{4\pi\Delta}{\delta} +
\frac{g_S^2}{4}
\end{eqnarray}
and
\begin{eqnarray} {\cal
K}=\frac{16\pi^2\Delta^2}{\delta^2}+\frac{(g_L+g_R)^2+g_S^2}{4}+g_S\frac{4\pi\Delta}{\delta}.
\end{eqnarray}

In the limit $\Delta\to 0$ our results for $G_{DET}$ and $G_{CAR}$ reduce 
to the corresponding expressions derived in Ref. \onlinecite{Morten} for the 
normal quantum dot. At the same time, our result (\ref{GDA}) for 
the Andreev conductance $G_{A}$ (for $\Delta\to 0$) 
turns out to be 4 times bigger than the analogous expression \cite{Morten}. 
This difference is supposed to be due to a different
definition of the Andreev conductance employed in Ref. \onlinecite{Morten}.    

Combining the above results for $G_{DET}$ and $G_{CAR}$ we immediately
arrive  at the zero temperature linear non-local conductance 
$G_{LR}= G_{DET}-G_{CAR}$ for our device. It reads
\begin{eqnarray}
G_{LR}= \frac{e^2}{16\pi}
\frac{g_Lg_R(g_L+g_R)^2}
{\left(
    \frac{16\pi^2\Delta^2}{\delta^2}+\frac{(g_L+g_R)^2+g_S^2}{4}+g_S\frac{4\pi\Delta}{\delta} \right)^{3/2}}.
\label{glrlinear}
\end{eqnarray}
This expression demonstrates again why the lowest order perturbation theory
in barrier transmissions \cite{Falci} yields zero non-local conductance at
$T=0$. This perturbation theory applies in the weak tunneling limit 
$g_{L,R} \ll 1$. The result (\ref{glrlinear}), however, contains only 
higher order terms in barrier transmissions whereas the contribution
$\propto g_Lg_R$ should vanish. 
This situation is qualitatively similar to that of NSN structures with
ballistic electrodes \cite{KZ06,KZ07}. We would also like to emphasize
that the exact cancellation of $G_{DET}$ and $G_{CAR}$ in the lowest
order in $g_Lg_R$ holds for any $g_S$ and {\it does not}
require taking the limit $g_S \to \infty$. This is in contrast to the case
of normal quantum dots \cite{Morten} in which $G_{LR}$ was found to vanish
{\it only} for  $g_S \to \infty$.

At small tunneling rates $\Gamma_L,\Gamma_R,\Gamma_S\ll\Delta$ 
Eq. (\ref{glrlinear}) reduces to
\begin{eqnarray}
G_{LR}=\frac{e^2}{(16\pi)^4}\frac{g_Lg_R(g_L+g_R)^2\delta^3}{\Delta^3}.
\label{ddd}
\end{eqnarray}
In the limit of a bulk metal $\delta\to 0$ (though $d \lesssim \xi_0$) 
the proximity effect becomes
unimportant and the non-local conductance $G_{LR}$ (\ref{ddd})
vanishes already to {\it all} orders in $g_{L,R}$.

Finally, we present the exact expression for the 
zero temperature non-local resistance $R_{LR}$. It reads
\begin{eqnarray}
\frac{R_{LR}}{R_S}=\frac{2\Gamma_S\left(\left(\Delta+\frac{\Gamma_S}{2}\right)^2
+\frac{(\Gamma_L+\Gamma_R)^2}{4}\right)^{3/2}}
{\left(2\left(\Delta+\frac{\Gamma_S}{2}\right)^2+\frac{(\Gamma_L+\Gamma_R)^2}{4}\right)^2
-(\Gamma_L+\Gamma_R)^2}.
\label{Rnl0}
\end{eqnarray}
In the limit $\Gamma_S\gg\Delta,\Gamma_L,\Gamma_R$ we get
$R_{LR}=R_S$, i.e. in this limit the non-local resistance just coincides with 
its normal state value. For $\Delta\gg \Gamma_L,\Gamma_R,\Gamma_S$
we obtain $R_{LR}=\delta/4e^2\Delta\ll R_S$.

\subsection{Non-zero temperatures}

Finally let us briefly discuss the effect of temperature on zero bias
conductances of our system. Combining Eqs. (\ref{nonlcurr})-(\ref{CARcurr})
and (\ref{cnds}) we obtain
\begin{eqnarray}
&& G_{LL}=
\frac{e^2\Gamma_L\Gamma_S}{\pi}\sum_n\int dE
\frac{|E|\theta(|E|-\Delta_S)}{\sqrt{E^2-\Delta_S^2}}
\nonumber\\ &&\times\,
\frac{1}{4T\cosh^2(E/2T)}\bigg\{
|G_R(E,\xi_n)|^2+|F_R(E,\xi_n)|^2
\nonumber\\ &&
-\,\frac{2\Delta_S}{E}\,{\rm Re}\,\big[G_R(E,\xi_n)F_R^*(E,\xi_n)\big]
\bigg\}
\nonumber\\ &&
+\, \frac{e^2\Gamma_L\Gamma_R}{\pi}\sum_n\int \frac{dE}{4T\cosh^2(E/2T)}
\nonumber\\ &&\times\,
\left\{|G_R(E,\xi_n)|^2+\left(\frac{2\Gamma_L}{\Gamma_R}+1\right)|F_R(E,\xi_n)|^2\right\},
\label{GLL0}
\end{eqnarray}
\begin{eqnarray}
G_{LR}=\frac{e^2\Gamma_L\Gamma_R}{\pi}\sum_n\int dE\,
\frac{ |G_R(E,\xi_n)|^2-|F_R(E,\xi_n)|^2}{4T\cosh^2(E/2T)}
\label{GLR0}
\end{eqnarray}
Substituting the expressions for the Green functions (\ref{GF0})-(\ref{PRA}) into
the above equations we arrive at the final results for zero-bias conductances
at non-zero $T$. These results are illustrated in Figs.4 and 5.

\begin{figure}
\begin{tabular}{cc}
(a) & \raisebox{-4.7cm}{\includegraphics[width=6.5cm]{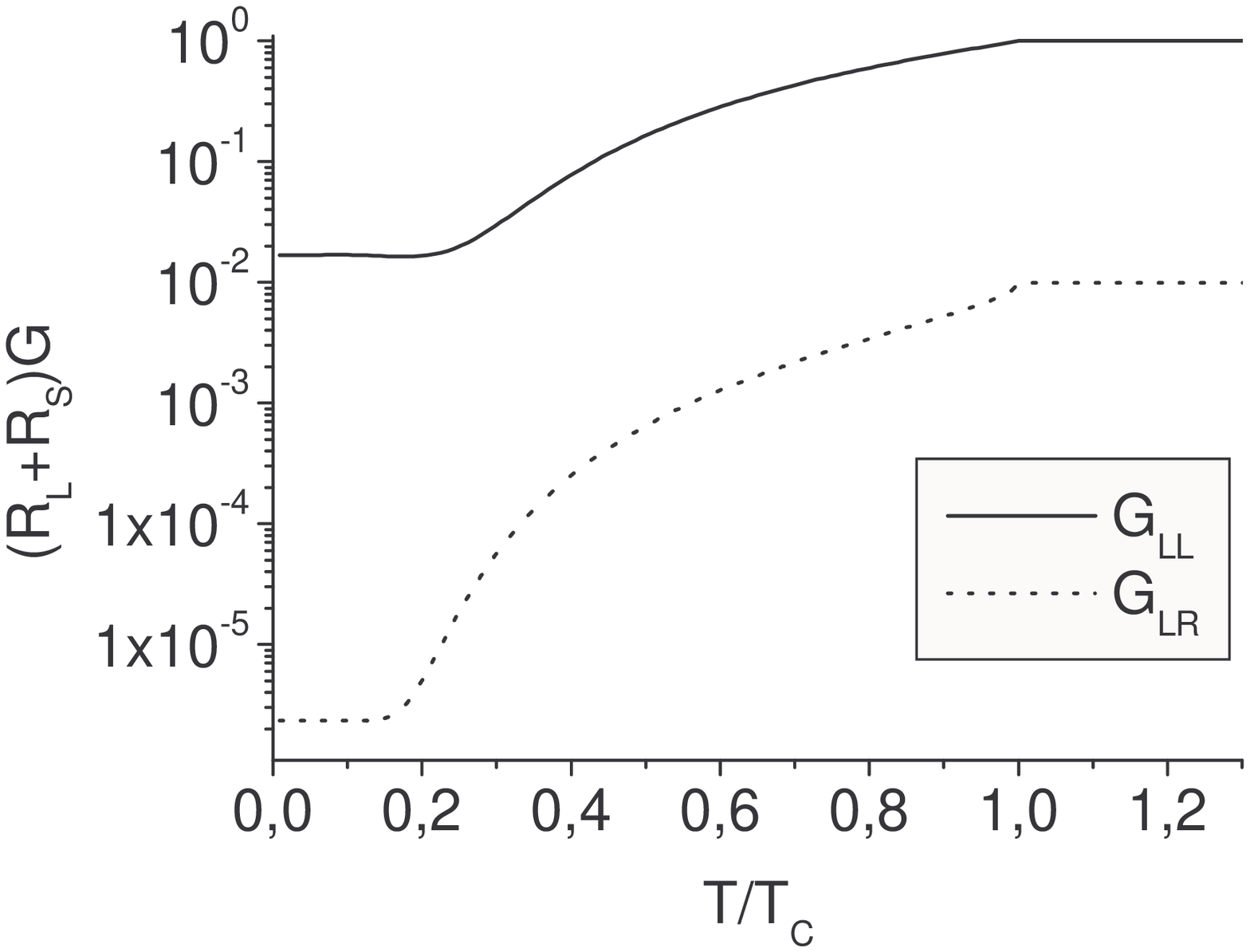}}
\\
(b) & \raisebox{-4.5cm}{\includegraphics[width=6cm]{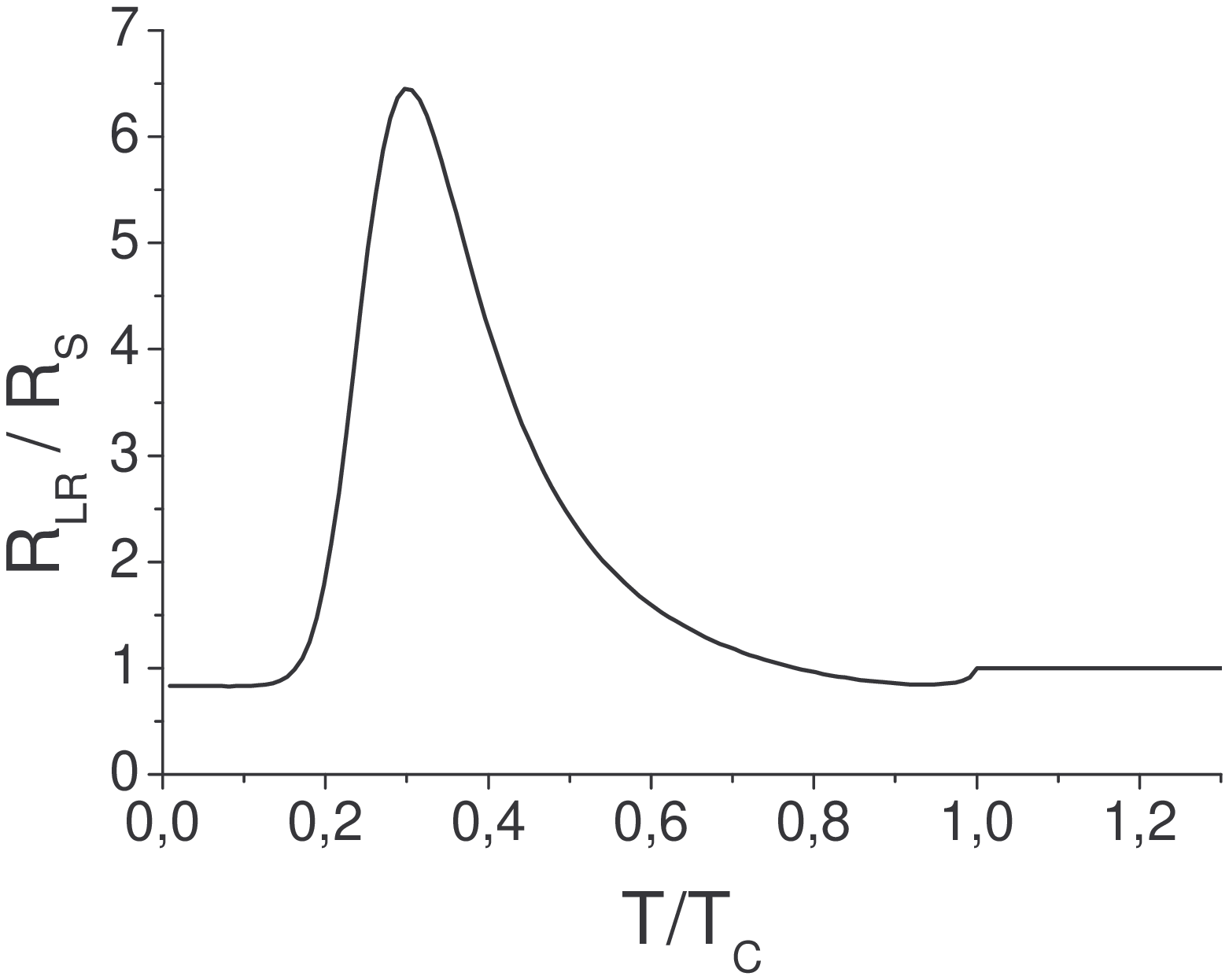}}
\end{tabular}
\caption{(a) temperature dependence of local $G_{LL}$ (\ref{GLL0}) and non-local 
$G_{LR}$ (\ref{GLR0}) zero-bias conductances. 
The parameters were chosen as follows: $\Delta=\Delta_S$, $\Gamma_L=\Gamma_R=0.1\Delta(0)$,
$\Gamma_S=10\Delta(0)$;
(b) non-local zero bias resistance $R_{LR}$ (\ref{Rnl}) normalized by its normal
state value (for the same parameters).}
\end{figure}

Fig. 4 shows the temperature dependence of both local and non-local zero
bias conductances (\ref{GLL0}), (\ref{GLR0}) along with that for the 
non-local resistance (\ref{Rnl}).
The conductances $G_{LL}$ and $G_{LR}$ decrease monotonously 
with decreasing temperature.
The temperature dependence of the non-local resistance $R_{LR}$ is, on the
contrary, non-monotonous.
At temperatures just below $T_C$ the resistance $R_{LR}$ first slightly
decreases but then it starts growing exponentially with decreasing $T$ due to 
charge imbalance effects, as it was explained in Sec. 3b. Such a tendency
persists down to the crossover temperature 
\begin{eqnarray}
T^* \sim \frac{\Delta}{\ln\frac{2\Delta +\Gamma_S}{\Gamma_L+\Gamma_R}},
\;\;\; \Gamma_L,\Gamma_R\ll 2\Delta+\Gamma_S
\label{Tstar}
\end{eqnarray}
at which the non-local resistance reaches its maximum.
Below  $T^*$,  AR 
contribution starts dominating over that caused by charge imbalance.
For this reason at $T < T^*$ $R_{LR}$ drops sharply and 
then at $T\sim T^*/2$ saturates to its zero temperature value (\ref{Rnl0}), 
as it is seen in Fig. 4. 

Note that qualitatively the same behavior of the non-local resistance 
was recently observed in experiments\cite{Beckmann,Venkat}, cf., eg., Fig. 3
in Ref. \onlinecite{Venkat}. Though a detailed quantitative comparison
between our theoretical predictions and the experimental results 
\cite{Beckmann,Venkat} is rather difficult to perform due to different geometry
of the model employed here, we believe that our theory correctly describes
the physical origin of the peak in the temperature dependence of the non-local
resistance observed in Refs. \onlinecite{Beckmann,Venkat}.

\begin{figure}
\includegraphics[width=6.5cm]{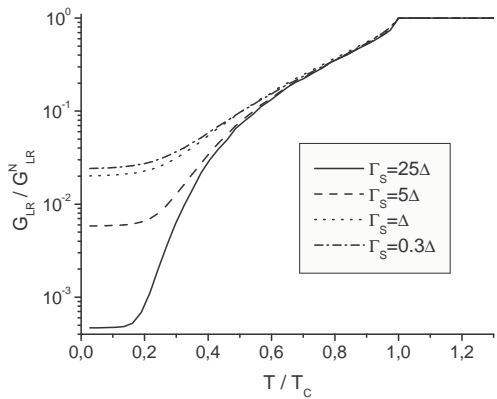}
\caption{Temperature dependence of zero-bias non-local conductance $G_{LR}$
for different values of $\Gamma_S$. Here we set $\Delta=\Delta_S$,
$\Gamma_L=\Gamma_R=0.3\Delta(0)$.}
\end{figure}

For completeness, in Fig. 5 we present the dependence $G_{LR}(T)$ at 
different values of $\Gamma_S$. As temperature decreases below
the critical temperature $T_C$ the conductance $G_{LR}(T)$ drops sharply 
below its normal state value (\ref{GN}) and at $T \ll T_C$ it 
saturates to the zero-temperature value which essentially depends of the
relation between $\Gamma_{L,R,S}$ and $\Delta$. We observe that for given 
 $\Gamma_{L,R}$ this value {\it decreases} with increasing coupling 
 $\Gamma_{S}$ between the dot and the superconducting lead. This tendency is
explained by the fact that CAR becomes progressively more pronounced with
increasing tunneling rate $\Gamma_{S}$.

\section{Conclusions}

In this paper we developed a  microscopic theory of  non-local electron 
transport in three-terminal NSN structures which consist of a superconducting 
chaotic
quantum dot (with typical size $d \lesssim \xi_0$) attached to one 
superconducting and two normal reservoirs, as it is shown in Fig. 1.
By varying the tunneling rates between the dot and the electrodes 
$\Gamma_{L,R,S}$ (which  play the role of effective Thouless energies
for electrons in the part of a superconducting electrode directly attached to
normal leads) one can cover a number of different physical situations and
limits and illustrate the relation to the models considered by other authors.

Our analysis is employed within the general Keldysh formalism which fully
accounts for non-equilibrium effects and disorder in the superconducting
terminal (dot). Our theory allows to go beyond perturbation theory in
dimensionless conductances between S- and N-electrodes $g_{L,R}$ and derive
a general expression for the conductance matrix which remains valid
in both weak and strong tunneling limits. This result enables one to study
and compare relative contributions to the non-local conductance provided
by the competing processes of direct electron transfer (DET) and crossed
Andreev reflection (CAR). We demonstrated that at low energies these
contributions do not cancel each other beyond the weak tunneling limit.
This is the result of the proximity effect: Coupling to normal electrodes 
yields non-zero subgap density of states inside the superconducting
dot which in turn causes a decrease of the CAR contribution to the
non-local conductance $G_{LR}$. On the contrary, increasing coupling
between the dot and the superconducting electrode increases CAR and, hence,
decreases  $G_{LR}$.

Our theory allows to investigate the effect of charge imbalance on
non-local electron transport in NSN devices. We argued that temperature
dependence of the non-local resistance $R_{LR}$ of such devices 
is determined by
the competition between charge imbalance and Andreev reflection.
The contribution of the former process dominates over
that of the latter at $T \gtrsim T^*$ (where $T^*$ is the crossover
temperature defined in Eq. (\ref{Tstar})) causing an increase
$R_{LR}(T)$ with decreasing $T$. In contrast, at lower temperatures 
AR dominates
and $R_{LR}(T)$ decreases as $T$ becomes lower. As a result, the 
dependence $R_{LR}(T)$ acquires a pronounced peak at 
$T \sim T^*$, see Fig. 4. This behavior was observed in recent
experiments \cite{Beckmann,Venkat}.

This work is part of the EU Framework Programme
NMP4-CT-2003-505457 ULTRA-1D "Experimental and theoretical investigation of
electron transport in ultra-narrow 1-dimensional nanostructures".

\end{document}